\documentclass[11pt]{article}
\begin{document}
\vspace*{1.2in}
\begin{center}\Large{\bf What is paradoxical about \\ 
the ``Three-box paradox''?}\\
\vspace{1cm}
\normalsize\ J. Finkelstein\footnote[1]{
        Participating Guest, Lawrence Berkeley National Laboratory\\
        \hspace*{\parindent}\hspace*{.5em}
        Electronic address: JLFINKELSTEIN@lbl.gov}\\
        Department of Physics\\
        San Jos\'{e} State University\\San Jos\'{e}, CA 95192, U.S.A
\end{center}
\begin{abstract}
This paper is a comment on quant-ph/0606067 by Ravon and Vaidman, in
which they defend the position that the ``three-box paradox'' is
indeed paradoxical.
\end{abstract}
\newpage
Some time ago, it was suggested (\cite{AAD}, \cite{AV}) that the
description of a quantum system in the time interval between two
measurements would involve aspects which are counter-intuitive and
perhaps even paradoxical.  One example which has been suggested involves 
a particle which might be in any of several boxes.  This example has
been given the suggestive title ``the three-box paradox'', but in
order to not pre-judge the issue I will refer to this example as the
``three-box story''.  
Other authors (\cite{BB}--\cite{Kirk})
have criticized various aspects of the interpretations given by the
authors of \cite{AAD} and  \cite{AV}.  In particular, it has been
argued that the three-box story should not be considered to be paradoxical,
but in a recent article  Ravon and Vaidman \cite{RV} have defended the
claim that this story  should indeed be called a paradox.

The three-box story concerns three boxes labeled $A$, $B$, and $C$,
and a single spinless particle.  Define $|A\rangle $ to be the state of
the particle if it is in box $A$, with analogous definitions for the states 
$|B\rangle $ and  $|C\rangle$.  The particle is prepared in the initial
state
\begin{equation}
|\Psi _{i}\rangle = \frac{1}{\sqrt{3}}(|A\rangle + |B\rangle +
 |C\rangle )
\end{equation}
and post-selected to be in the state
\begin{equation}
|\Psi _{f}\rangle = \frac{1}{\sqrt{3}}(|A\rangle + |B\rangle -
 |C\rangle )
\end{equation}
At some time after the preparation but before the post-selection an
agent, let me call her Alice, looks into either box $A$ or box $B$ to
see whether the particle is there; an elementary calculation shows
that Alice surely finds the particle in whichever of these two boxes
she happens to look.

Kirkpatrick \cite{Kirk} has described a purely-classical situation
which parallels that in the three-box story.  The obvious implication
of this would be  that since a classical story would not be paradoxical,
the three-box story must also not be.  Ravon and Vaidman dispute this
implication by claiming that Kirkpatrick's story and the three-box
story are not really analogous.  Of course, since the three-box story
includes quantum measurements, no classical story could reproduce it
exactly; the question to be decided is: does Kirkpatrick's story
capture those aspects of the three-box story which are asserted 
to be paradoxical?

So what is it about the three-box story which is asserted to be paradoxical?
Here are two quotations from the paper by Ravon and Vaidman \cite{RV};
I will label these quotations RV1 and RV2. 
Near the bottom of page 1, Ravon and Vaidman write
\begin{description}
\item[RV1] The paradox in the Three-Box experiment is that at a
  particular time we can claim that a particle is in some sense both
  with certainty in one box, $A$, and with certainty in another box,
  $B$.  Now, if a particle is certainly in $A$, then it is certainly
  not in $B$, and vice versa.  Therefore, if a single particle is
  both certainly in $A$ and certainly in $B$ we have a paradox.
\end{description}
Then on page 2 they write
\begin{description}
\item[RV2] In the Three-Box experiment, the particle is {\em certain}
  to be found in $A$ if searched for in $A$, and {\em certain} to be
  found in $B$ if searched for in $B$ instead.
\end{description}

Before agreeing that quotation RV2 does indeed describe a paradox, we
should remember the condition under which it is supposed to be true,
namely that the particle was post-selected in the state 
$|\Psi _{f}\rangle $ which was given in eq.\ 2.  
Let me say that the post-selection is accomplished by
another agent (whom I will call Bob) measuring an observable (which I
will call $R$) whose associated operator has as eigensubspaces the
subspace generated by $|\Psi _{f}\rangle $ and the complement of that
subspace.  If when he measures $R$, Bob does indeed find the state
$|\Psi _{f}\rangle $, I will say that the post-selection ``succeeds''.
Then by including together with RV2 the explicit mention of the
condition under which it is supposed to be valid, we get a statement I
will label S1:
\begin{description}
\item[S1] If the post-selection succeeds, then
\begin{enumerate}
  \item If the particle was searched for in box $A$, it was certainly
  found there, and
  \item If the particle was searched for in box $B$, it was certainly
  found there.
\end{enumerate} \end{description}  

Statement S1 has the same content as does quotation RV2 (with  the
condition which was implicit there made explicit), but Kirkpatrick
could make an equivalent statement about his classical story (after
correcting for the fact that his story involves a deck of
cards rather than a particle and boxes \cite{Kirkq}). Nevertheless,
Ravon and Vaidman insist that the three-box story is paradoxical in a
way in which Kirkpatrick's story is not; they define a ``quantum
paradox'' to be ``a phenomenon that {\em classical} physics cannot
explain''.  Obviously that is a definition which no classical story
can meet, but I would suggest that it   does not represent what is
usually understood by the term ``paradox''.  Suppose that quantum
physics permitted the single particle to suddenly become three particles,
one in each box; that might not be explained by classical physics, but
would not usually be called paradoxical.  A ``paradox'' is usually
considered to be something which (at least
apparently) violates the laws of common sense, or perhaps even those
of logic. And that is the way in which
the advocates of the ``three-box paradox'' advertise it (at least when
they are not arguing against Kirkpatrick's example); quotation RV1   
clearly claims that the three-box story is in (at least apparent)
conflict with laws of
logic. Furthermore, any quantum measurement will have non-classical
aspects, so if we were to accept the Ravon-Vaidman definition of
quantum paradox, we might conclude that the three-box story does
involve a paradox, but only to the extent that any quantum measurement
does.  I suspect that Ravon and Vaidman would want us to conclude more
than that.

Kastner \cite{Kast} has re-told the three-box story following the
ordinary temporal order: First, the particle is prepared in the state
$|\Psi _{i}\rangle $ given in eq.\ 1.  Next, Alice looks into either
box $A$ or box $B$; either way she might find the particle (the
probability that she finds it is
1/3) or she might not find it.  Finally, Bob measures
the observable $R$; in the cases in   which Alice did find the
particle in the box into which she looked, this post-selection might
succeed (also with probability 1/3, as it happens), but in the cases in
which Alice did not find the particle, the post-selection
certainly does not succeed.  Thus we can make a statement which I 
label S2:
 \begin{description}
\item[S2] 
\begin{enumerate}
  \item If the particle was searched for in box $A$ and not found
  there,\\  then the post-selection does not succeed, and
  \item If the particle was searched for in box $B$ and not found
  there, then the post-selection does not succeed.
\end{enumerate} \end{description} 

According to statement S2, two different and incompatible scenarios
(namely, particle searched for in box $A$ but not found, and particle
searched for in box $B$ but not found) have a common consequence
(namely, the post-selection does not succeed), but there is nothing at
all strange about that.  The calculation which leads to S2 does
involve quantum interference, but quantum interference by itself is
not usually
considered paradoxical (at least, not now in the 21st century!) 
In fact, there does not seem to be any reason
whatsoever to label as paradoxical the situation described by
statement S2.  But statements S1 and S2 are completely equivalent, so
if S2 is not paradoxical, then neither is S1. 

Quotation RV2 is essentially the same as statement S1, except that the
restriction to cases in which the post-selection succeeds is implicit
in RV2
(it is part of the definition of the three-box experiment) while in S1
that restriction is stated explicitly. Certainly Ravon and Vaidman
never state, nor in any way imply, that RV2 would be valid without
that restriction---if they had, their paper would simply be wrong, and
that is not at all the case.  Why then does RV2 at least give the
impression of describing a paradox?  What one judges to be paradoxical is
to some extent a matter of personal psychology, but let me speculate
that when the restriction to cases in which the post-selection
succeeds is not made explicitly, the reader might tend to not consider
the  significance of that restriction.  The reader might forget 
that there is never a situation in which the post-selection has already
succeeded but for which Alice's looking into boxes $A$ and $B$ are
both still possible. (Or, as pointed out  by Bub and
Brown \cite{BB}, that the post-selected  subensemble in which Alice  
looks into box
$A$ does not coincide with that in which she looks into box $B$.)
In any event, when the three-box story is summarized as in  
statement S1, or even better, as in  statement S2, there does not
seem to be any reason to consider it paradoxical.

Statement S2 completely describes all of the relevant features of the
three-box story, and shows that it is about as straightforward as any
story involving quantum interference could be.  Does this not indicate
that any other account of this story which suggests that it might have
paradoxical aspects must be misleading?

\vspace{1cm}
Acknowledgement: I would  like to acknowledge the hospitality of the
Lawrence Berkeley National Laboratory, where this work was done.


\begin{thebibliography}{99}
\bibitem{Kirkq}In parallel   with statement S1, Kirkpatrick could say
  of his example ``If the post-selection for Face = K succeeds, then
  1. If Suit was checked to be S, it was certainly found to be so, and
  2. If Suit was checked to be D, it was certainly found to be so.''
\end{thebibliography}
\end{document}